\documentstyle[epsfig]{aipproc}
\begin{document}

\title{Color Confinement in QCD\\ due to Topological Defects}

\author{Kei-Ichi Kondo$^*$}
\address{$^*$Department of Physics, Faculty of Science, 
Chiba
University, Chiba 263-8522, Japan
\thanks{Talk presented at {\it New Directions in Quantum
Chromodynamics}, June 21-25, 1999, Kyungju, Korea.  This work is
supported in part by the Grant-in-Aid for Scientific Research from
the Ministry of Education, Science and Culture (No.10640249). }\\
%Boulder Colorado 80307\\
%$^{\dagger}$National Standards Institute, Boulder, Colorado 11543
}

%\lefthead{LEFT head}
%\righthead{RIGHT head}
\maketitle

\begin{abstract}
We outline a derivation of area law of the Wilson loop in SU(N) 
Yang-Mills theory in the maximal Abelian gauge.  This is based on a
new version of non-Abelian Stokes theorem and the novel
reformulation of the Yang-Mills theory.  Abelian dominance and
monopole dominance of the string tension in SU(N) QCD are immediate
consequences of this derivation.
\end{abstract}

\section*{Introduction}

\par
In the dual superconductor picture \cite{DSC} of quark confinement,
the magnetic monopoles give the dominant contribution to the area
law of the Wilson loop or the string tension.      
Based on 't Hooft argument \cite{tHooft81}, the partial gauge
fixing $G \rightarrow H$ realizes the magnetic monopole in
Yang-Mills gauge theory in the absence of elementary scalar field.
In the conventional approach based on the maximal Abelian (MA)
gauge, the residual gauge group was chosen to be the maximal torus
subgroup
$H=U(1)^{N-1}$ for $G=SU(N)$.  This choice immediately determines the
type of magnetic monopoles.  
We re-examine this issue.  We learn that the magnetic monopole which
is responsible for area law of the Wilson loop is determined by
the maximal stability group $\tilde H$ rather than the residual gauge
group $H$. This is a new feature appeared in 
$SU(N), N \ge 3$, which was overlooked so far in the lattice
community as far as I know. Indeed, this situation occurs only for
$SU(N), N \ge 3$.  Therefore, we must distinguish the maximal
stability group
$\tilde H$  from the residual gauge group $H$.
In general, the maximal stability group $\tilde H$ is larger than
the maximal torus subgroup, 
$H=U(1)^{N-1} \subset \tilde H$. So, the coset
$G/\tilde H \subset G/H$.
We derive area law of the Wilson loop in $SU(N)$
Yang-Mills theory in the MA gauge.  This is
performed based on the results of a series of works
\cite{KondoI,KondoII,KondoIII,KondoIV,KondoV,KondoVI}.

\section*{Coherent state}
\par
First of all, we construct the coherent state
$
  |\xi, \Lambda \rangle 
$
corresponding to the
coset representatives $\xi \in G/\tilde H$.  
We define the maximal stability subgroup (isotropy subgroup) $\tilde
H$ as a subgroup of
$G$ that consists of all the group elements $h$ that leave the
highest-weight state $|\Lambda \rangle$ invariant up to a phase
factor, i.e.,
$
  h |\Lambda \rangle = |\Lambda \rangle e^{i\phi(h)}, h \in \tilde H .
$
The phase factor is unimportant in the following discussion because
we consider the expectation value of any operators in the coherent
state.  The maximal stability subgroup $\tilde H$ includes the Cartan
subgroup
$H=U(1)^r$, i.e.,
$H \subset \tilde H$.
\par
Let $H$ be the Cartan subgroup of $G$ and ${\cal H}$ be the Cartan
subalgebra in ${\cal G}$.
For every element
$g\in G$, there is a unique decomposition of
$g$ into a product of two group elements, 
$
 g = \xi h, \xi \in G/\tilde H, h \in \tilde H ,
$
for $g \in G$.
We can obtain a unique coset space $G/\tilde H$ for a given $|\Lambda \rangle$.
The action of arbitrary group element $g\in G$  on 
$|\Lambda \rangle$ is given by
$
  g |\Lambda \rangle = \xi h  |\Lambda \rangle
  = \xi  |\Lambda \rangle  e^{i\phi(h)}.
$
\par
The coherent state is constructed as
$
  |\xi, \Lambda \rangle = \xi |\Lambda \rangle .
$
This definition of the coherent state is in one-to-one
correspondence with the coset space $G/\tilde H$ and the coherent states
preserve all the algebraic and topological properties of the coset
space $G/\tilde H$.
If $\Gamma^\Lambda({\cal G})$ is Hermitian, then $H_i^\dagger=H_i$, 
and $E_{\alpha}^\dagger = E_{-\alpha}$.  
Every group element $g \in G$ can be written as  the exponential of
a complex linear combination of diagonal operators $H_i$ and
off-diagonal shift operators $E_\alpha$.  Thus the coherent
state is given by 
\begin{eqnarray}
  |\xi, \Lambda \rangle 
  = \xi |\Lambda \rangle
  = \exp \left[ 
  \sum_{\beta '} \left( \eta_\beta E_\beta - \bar{\eta}_\beta E_{-\beta}
\right) \right] |\Lambda \rangle,
\quad \eta_\beta \in \bf{C},
\end{eqnarray}
where 
$|\Lambda \rangle$ is the highest-weight state 
($H_j | \Lambda \rangle = \Lambda_j | \Lambda \rangle$, 
$E_\alpha | \Lambda \rangle = 0$ for $\alpha \in R_+$, $R_+ (R_-)$
is a subsystem of positive (negative) roots.)  such that 
\begin{enumerate}
\item[(i)]
 $|\Lambda \rangle$  is
annihilated by all the (off-diagonal) shift-up operators $E_{\alpha}$ with
$\alpha>0$, 
$
 E_{\alpha} |\Lambda \rangle = 0 (\alpha>0) ;
$
\item[(ii)] 
$|\Lambda \rangle$  is
mapped into itself by all diagonal operators $H_i$,
$
 H_{i} |\Lambda \rangle = \Lambda_i |\Lambda \rangle ;
$
\item[(iii)] 
$|\Lambda \rangle$  is
annihilated by some shift-down operators $E_{\alpha}$ with
$\alpha<0$, not by other $E_{\beta}$ with $\beta<0$:
$
 E_{\alpha} |\Lambda \rangle = 0 ({\rm some~} \alpha<0) ;
$
$
 E_{\beta} |\Lambda \rangle = |\Lambda+\beta \rangle 
 ({\rm some~} \beta<0) ;
$
and the sum $\sum_{\beta '}$ is restricted to those shift operators
$E_{\beta}$ which obey (iii).
\end{enumerate}
\par
The coherent states are normalized to unity, 
$
 \langle \xi, \Lambda | \xi, \Lambda \rangle = 1 ,
$
but, are non-orthogonal, 
$
 \langle \xi' , \Lambda | \xi, \Lambda \rangle \not= 0 .
$
The coherent state spans the entire space $V^\Lambda$.
By making use of the the group-invariant measure $d\mu(\xi)$ of
$G$ which is appropriately normalized, we obtain
$
  \int |\xi, \Lambda \rangle d\mu(\xi) 
  \langle \xi,  \Lambda |
  = I ,
$
which shows that the coherent states are complete, but overcomplete.
This resolution of identity is very important to obtain the path
integral formula given below.
The coherent states 
$|\xi, \Lambda \rangle$
are in one-to-one correspondence with the coset representatives $\xi \in G/\tilde H$,
$
 |\xi, \Lambda \rangle \leftrightarrow G/\tilde H .
$
In other  words, $|\xi, \Lambda \rangle$
and $\xi \in G/\tilde H$ are topologically equivalent.
\par
For concreteness, we first focus on the SU(3) case.  
The highest weight $\Lambda$ of the representation specified
by the Dynkin index $[m,n]$ ($m,n$: integers) can be written as
$
  \vec \Lambda = m \vec h_1 + n \vec h_2
$
($m,n$ are non-negative integers for the highest weight)
where $h_1, h_2$ are highest weights of two fundamental
representations of SU(3) corresponding to $[1,0], [0,1]$
respectively, i.e.,
$
  \vec h_1 = \left({1 \over 2}, {1 \over 2\sqrt{3}} \right), \quad
  \vec h_2 = \left({1 \over 2}, {-1 \over 2\sqrt{3}} \right) .
$
Therefore, we obtain
$
  \vec \Lambda =  \left({m+n \over 2}, {m-n \over 2\sqrt{3}} \right)
.
$
If $mn=0$, i.e.,
$m=0$ or $n=0$, the maximal stability group $\tilde H$ is given by 
$\tilde H=U(2)$  (case (I)). 
Such a degenerate case occurs when the highest-weight vector $\vec
\Lambda$ is orthogonal to some root vectors. 
If $mn \not=0$, i.e.,
$m\not=0$ and $n\not=0$, $H$ is the maximal torus group 
$\tilde H=U(1) \times U(1)$ (case (II)). This is a non-degenerate
case. Therefore, for the highest weight
$\Lambda$ in the case (I), the coset $G/\tilde H$ is given by
\begin{equation}
  SU(3)/U(2)=SU(3)/(SU(2)\times U(1))=CP^2,
\end{equation}
whereas in the case (II) 
\begin{equation}
  SU(3)/(U(1)\times U(1))=F_2 .
\end{equation}
Here, $CP^{n}$ is the
complex projective space and $F_n$ is the flag space. Therefore, the
two fundamental representations belong to the case (I), so the
maximal stability group is $U(2)$, rather than the maximal torus
group $U(1) \times U(1)$.  

\section*{Non-Abelian Stokes theorem}
We have derived a new version of non-Abelian Stokes theorem (NAST)
\cite{KT99a,KT99b}. For the non-Abelian Wilson loop defined by the
trace of the path-ordered exponent along the closed loop $C$,
\begin{equation}
  W^C [{\cal A}] :=  {\rm tr} \left[ {\cal P} 
 \exp \left( i g \oint_C {\cal A}
 \right) \right] ,
\end{equation}
with ${\cal A}$ being the connection one-form, 
$
 {\cal A}(x) = {\cal A}_\mu^A(x)T^A dx^\mu = {\cal A}^A(x) T^A ,
$
the NAST for $SU(N)$ is given by
\begin{eqnarray}
 W^C[{\cal A}] 
&=& \int [d\mu(\xi)]_C \exp \left( 
i g \oint_C  \left[ n^A {\cal A}^A  
+ {1 \over g} \omega \right] \right)  
\nonumber\\
&=& \int [d\mu(\xi)]_C \exp \left( 
ig \int_{S:\partial S=C} 
\left[ d(n^A {\cal A}^A) + {1 \over g} \Omega_K \right] \right) ,
\label{NAST}
\end{eqnarray}
where we have defined
\begin{eqnarray}
 n^A(x) &:=&
 \langle \Lambda | V(x) T^A V^\dagger(x) |\Lambda \rangle ,
\\
  \omega(x) 
 &:=& \langle \Lambda | i V(x) d V^\dagger(x) |\Lambda \rangle  
  = \langle \Lambda | i \xi^\dagger (x) d\xi(x) |\Lambda
\rangle ,
\end{eqnarray}
and $\Omega_K$ is the K\"ahler two-form given by
\begin{equation}
 \Omega_K := d\omega.
\end{equation}
The NAST (\ref{NAST}) implies that  the
Wilson loop  is rewritten into
\begin{eqnarray}
 W^C[{\cal A}]  
 =    \int [d\mu(\xi)]_C  \exp \left( i g \oint_{C} a \right)
 =   \int [d\mu(\xi)]_C  \exp \left( 
i g \int_{S:C=\partial S} f \right) .
\label{NAST2}
\end{eqnarray}
First, $a$ is the connection one-form,
\begin{equation}
 a := n^A {\cal A}^A  + {1 \over g} \omega 
 = \langle \Lambda | {\cal A}^V |\Lambda \rangle ,
\end{equation}
where
$
{\cal A}^V:=V{\cal A} V^\dagger + {i \over g}V d V^\dagger 
$
is the gauge transformation of ${\cal A}$ by
$V \in F_{N-1}$.  For quark in the fundamental representation,
\begin{equation}
 a  = {\rm tr}({\cal H}{\cal A}^V) .
\end{equation}
Therefore, the one-form $a$ is equal to the diagonal
piece of the gauge-transformed potential ${\cal A}^V$.
Next, $f$ is the curvature two-form,
\begin{eqnarray}
f := da 
= dC +  {1 \over g} d\omega 
= dC + {1 \over g} \Omega_K ,
 \label{apf}
\end{eqnarray}
where we defined the one-form,
$
  C := n^A {\cal A}^A .
$
The first piece $dC$ in $f$ does not contribute to the magnetic
current, due to the Bianchi identity.
On the other hand, the second term $\Omega_K$ in $f$
leads to the non-vanishing magnetic current,
\begin{equation}
  k_\mu := \partial_\nu {}^*f_{\mu\nu} \not= 0 ,
\end{equation}
where ${}^*f_{\mu\nu}$ is the Hodge dual of $f_{\mu\nu}$ in four
dimensions, 
$
   {}^*f_{\mu\nu}  := {1 \over 2} \epsilon_{\mu\nu\rho\sigma}
  f_{\rho\sigma} .
$
In general, the (curvature)
two-form 
$f=d(n^A {\cal A}^A) + \Omega_K$ in the NAST is the Abelian field
strength (which is invariant even under the non-Abelian  gauge
transformation of $G=SU(N)$), i.e., the generalized 't
Hooft-Polyakov tensor for $SU(N)$,
\begin{equation}
 f_{\mu\nu}(x) :=  \partial_\mu(n^A(x){\cal A}_\nu^A(x)) 
   - \partial_\nu(n^A(x){\cal A}_\mu^A(x))
   + {i \over g}{\bf n}(x) \cdot [\partial_\mu {\bf n}(x),
\partial_\nu {\bf n}(x)] .
\label{tHPtensor}
\end{equation} 
The invariance of $f$ is obvious from the NAST (\ref{NAST2}), since
the L.H.S. of (\ref{NAST2}), i.e., $W^C[{\cal A}]$ is gauge invariant
and the measure 
$[d\mu(\xi)]_C$ in the R.H.S. is also invariant under the $G$ gauge
transformation. Otherwise, the R.H.S. is zero.
In the case of fundamental representation,  the invariance is easily
seen, because  it is possible to rewrite  (\ref{tHPtensor})
into the manifestly gauge invariant form:
$
 f_{\mu\nu}(x) :=  {\rm tr}\left({\bf n}(x) {\cal F}_{\mu\nu}(x)
   + {i \over g}{\bf n}(x) \cdot
 [D_\mu {\bf n}(x), D_\nu {\bf n}(x)] \right) ,
$
where
$
{\cal F}_{\mu\nu}(x)
   := \partial_\mu {\cal A}_\nu(x) - \partial_\nu {\cal A}_\mu(x)
   - ig [{\cal A}_\mu(x), {\cal A}_\nu(x)] ,
$
and
$
  D_\mu {\bf n}(x) := \partial_\mu  {\bf n}(x)
- i g [{\cal A}_\mu(x), {\bf n}(x)] .
$

\section*{Abelian dominance}

\par
In our framework, the Abelian dominance and the monopole dominance
are understood as implying the first and the second equality
respectively,
\begin{eqnarray}
  \Big\langle W^C[{\cal A}]  \Big\rangle_{YM} 
   \cong  \Big\langle \exp \left( i g \oint_{C}  a \right)
\Big\rangle_{APEGT} 
 \cong  \Big\langle \exp \left( i \oint_{C}  \omega \right)
\Big\rangle_{APEGT} ,
\end{eqnarray}
where APEGT denotes the Abelian-projected effective gauge theory
\cite{KondoI}.
 Numerical simulations show that the monopole part
exhibits the area law and $\sigma_{Abel}$ exhausts the full string
tension obtained from  the non-Abelian Wilson loop (i.e., monopole
dominance in the string tension or area law),
\begin{eqnarray}
  \Big\langle \exp \left( i \oint_{C}  \omega \right)
\Big\rangle_{APEGT} 
 \sim& \exp (- \sigma_{Abel} |S|) ,
\end{eqnarray}
while 
$
\Big\langle \exp \left( i g \oint_{C}  a - i \oint_C \omega \right)
\Big\rangle_{APEGT}
$
does not exhibits the area law.
This result implies that the area law of the original non-Abelian
Wilson loop,
\begin{eqnarray}
  \Big\langle W^C[{\cal A}]  \Big\rangle_{YM}  
 \sim& \exp (- \sigma |S|) , \quad \sigma \cong \sigma_{Abel} .
\end{eqnarray}
The monopole dominance in this sense was derived for $SU(2)$ in
\cite{KondoIV} by showing that the dominant contribution to the
area law comes from the monopole piece alone,
$
 \Omega_K  = d\omega .
$
In \cite{KondoV}, the monopole dominance and the area law of the
Wilson loop have been shown based on the APEGT for $G=SU(2)$.
Now this scenario can be extended into $G=SU(N)$.

\section*{Magnetic monopoles in Yang-Mills theory}

\par
 The existence of magnetic
monopole is suggested from the non-trivial Homotopy groups
$\pi_2(G/H)$. In the case (II),
$
 \pi_2(F_2) = \pi_2(SU(3)/(U(1) \times U(1)))  
=\pi_1(U(1) \times U(1)) ={\bf Z}+{\bf Z} .
$
On the other hand, in the case (I), i.e., [m,0] or [0,n]
$
 \pi_2(CP^2) = \pi_2(SU(3)/U(2)) = \pi_1(U(2)) 
= \pi_1(SU(2)\times U(1)) = \pi_1(U(1))={\bf Z} .
$
Note that $CP^n$ NLSM has only the local $U(1)$ invariance for any
$n$.  It is this U(1) invariance that corresponds to a kind of
Abelian magnetic monopole in the case (I).
This magnetic monopole may be related to the non-Abelian magnetic
monopole proposed by E. Weinberg et al.
\par
This situation should be compared with the $SU(2)$ case where the
maximal stability group is always given by the maximal torus $H=U(1)$
irrespective of the representation.  Therefore, the coset is given by
$
G/H=SU(2)/U(1)=F_1=CP^1 \cong  S^2  \cong SO(3)
$
and 
$
 \pi_2(SU(2)/U(1)) = \pi_2(F_1) = \pi_2(CP^1) = {\bf Z} ,
$
 for {\it arbitrary} representation .
Actually, the NAST derived in this paper
shows that the fundamental quark feels only the $U(1)$ embedded in
the maximal stability group
$U(2)$ as a magnetic monopole (This is a component along the
highest-weight).
\par

\section*{Area law of the Wilson loop}
The (full) string tension $\sigma$ is defined by
\begin{eqnarray}
  \sigma := - \lim_{A(C) \rightarrow \infty}
  {1 \over A(C)} \ln \langle W^C[{\cal A}] \rangle_{YM_4} ,
\end{eqnarray}
where $A(C)$ is the minimal area spanned by the Wilson loop $C$. 
\par
We estimate the Wilson loop in the reformulation of the Yang-Mills
theory which was proposed by the author and was called the
perturbative deformation of a topological quantum field theory
\cite{KondoII}.  The Wilson loop is written as
\begin{eqnarray}
 \langle W^C[{\cal A}] \rangle_{YM_4} 
= \Biggr\langle \Biggr\langle 
  \exp \left[ i g \oint_C dx^\mu n^A(x) {\cal V}_\mu^A(x) \right]
\Biggr\rangle_{pYM_4}
  \exp \left[ i \oint_C \omega \right]
\Biggr\rangle_{TQFT_4} .
\end{eqnarray}
Here the expectation value 
$
 \langle  (\cdots) \rangle_{pYM}^{{\cal V}}
$
for the field ${\cal V}$ is calculated in the perturbation theory in
the coupling constant $g$.  
On the other hand, the expectation value
$
 \langle  (\cdots) \rangle_{TQFT}^U 
$
should be calculated in a non-perturbative way to incorporate the
topological contribution where $U$ is a compact gauge variable,
\begin{eqnarray}
  S_{TQFT}[\Omega, C, \bar C, B]
  := \int_{{\bf R}^4} d^4x \ i \delta_B \bar \delta_B 
  {\rm tr}_{G/H} \left[ {1 \over 2} \Omega_\mu(x) \Omega_\mu(x)
  + i C(x) \bar C(x) \right] ,
  \label{TQFT}
\end{eqnarray}
where 
$
 \Omega_\mu(x) := {i \over g} U(x) \partial_\mu U^\dagger(x).
$
This reformulation leads to the result:
\begin{eqnarray}
 \langle W^C[{\cal A}] \rangle_{YM_4} 
 = \Biggr\langle  \exp \left[ i \oint_C \omega \right]
 \Biggr\rangle_{TQFT_4} 
  \left[ 1  + O(g^2)  \right] .
\end{eqnarray}
This implies the magnetic monopole dominance in the area law of the
Wilson loop.

%\section*{Area law of the Wilson loop}

 For the planar Wilson loop $C$, the Parisi-Soulous dimensional
reduction \cite{KondoII} leads to
\begin{equation}
 \Big\langle \exp \left[ i \oint_C \omega \right]
\Big\rangle_{TQFT_4} 
=  \Big\langle \exp \left[ i \oint_C \omega \right]  
\Big\rangle_{NLSM_2} ,
\end{equation}
where the
two-dimensional nonlinear sigma model (NLSM) has the action,
\begin{eqnarray}
  S_{NLSM}[U, C, \bar C]
  := 2\pi \int_{{\bf R}^{2}} d^{2}x    \
  {\rm tr}_{G/H} \left[ {1 \over 2} \Omega_\mu(x) \Omega_\mu(x)
  + i C(x) \bar C(x) \right] .
  \label{GF'}
\end{eqnarray}
By making use of the complex coordinates of the flag space $G/H$,
the action is rewritten as 
\begin{eqnarray}
  S_{NLSM} &=& {\pi \over g^2} \int_{{\bf R}^{2}} d^{2}x
g_{\alpha\bar \beta}
  {\partial w^{\alpha} \over
\partial x_a}{\partial \bar w^{\beta} \over \partial x_a} 
\quad (a=1, 2) 
\\
&=& {\pi \over g^2} \int_{{\bf C}} dzd\bar z \
g_{\alpha\bar
\beta} 
 \left(
 {\partial w^\alpha \over \partial z}{\partial \bar w^\beta \over
\partial \bar z}
  + {\partial w^\alpha \over \partial \bar z}{\partial \bar w^\beta
\over \partial z}
  \right) ,
\end{eqnarray}
where $z=x+iy=x_1+ix_2 \in  {\bf C} \cong {\bf R}^2$,
and
$dxdy=dx_1 dx_2={i \over 2}dz d\bar z$,
and we have omitted the ghost term, $C(x) \bar C(x)$.
Here $g(\mu)$ is the
running Yang-Mills coupling constant whose running is
given by the perturbative deformation in four-dimensional
Yang-Mills theory. 
For the quark in the fundamental representation of $SU(N)$, the
relevant NLSM is given by $CP^{N-1}$ model. 
We can show the area law,
\begin{eqnarray}
  \Biggr\langle  \exp \left[ i \oint_C \omega \right]
\Biggr\rangle_{CP^{N-1}_2}
\sim \exp (- \sigma_0 TR ) ,
\end{eqnarray}
by the instanton calculus (dilute
instanton-gas approximation) or by the large $N$ expansion in the
two-dimensional
$CP^{N-1}$ model, see
\cite{KT99b}.
\par
In summary, we have given a new derivation of non-Abelian Stokes
theorem for
$G=SU(N)$ for $N\ge 2$ which reduces to the previous result for
$SU(2)$.  
This version of non-Abelian Wilson loop is very helpful to see the
role played by the magnetic monopole in the calculation of the
non-Abelian Wilson loop.
Combining this non-Abelian Stokes theorem
with the Abelian-projected effective gauge theory for
$SU(N)$, we have explained the Abelian dominance for the Wilson loop
in $SU(N)$ Yang-Mills gauge theory.
For $SU(N)$ with $N$ greater than two, we must distinguish the
maximal stability group $\tilde H$ and the residual gauge group 
$H=U(1)^{N-1}$.
\par
By making use of the non-Abelian Stokes theorem  in a novel
reformulation of the Yang-Mills theory proposed by
one of the authors
\cite{KondoII}, the derivation of the area law of the
non-Abelian Wilson loop in four-dimensional Yang-Mills theory has
been reduced to the two-dimensional problem of calculating the
expectation value of the Abelian Wilson loop in the coset
$G/H$ non-linear sigma model.
Especially, in order to show confinement of the fundamental quark in
four-dimensional $SU(N)$ Yang-Mills theory in the MA gauge, we have
only to consider the two-dimensional $CP^{N-1}$ model.
The details will be given in \cite{KT99a,KT99b}.
A Monte Carlo simulation on a lattice
will be efficient to confirm the above 
picture of quark confinement.

\end{document}